\documentclass{article}
\usepackage{spconf,amsmath,graphicx}
\usepackage[dvipsnames]{xcolor}
\usepackage[]{pgf}
\usepackage{tikz}
\usetikzlibrary{shapes,arrows,snakes,backgrounds,matrix,patterns,positioning,fadings}
\usepackage[mode=buildnew]{standalone}
\usepackage{subfig}
\usepackage{upgreek}
\usepackage{nicefrac}
\usepackage{dsfont}
\usepackage{bm}
\usepackage{cancel}
\usepackage{amsbsy}
\usepackage{amssymb}
\usepackage[ruled]{algorithm2e}
\usepackage{lipsum}
\usepackage{harpoon}
\usepackage[percent]{overpic}
\usepackage{comment}

\usepackage{hyperref}
\hypersetup{
	colorlinks,
	linkcolor={blue!80!black},
	citecolor={blue!80!black},
	urlcolor={blue!80!black}
}
\usepackage[short]{optidef}

\newcommand{\mtxb}[1]{\bm{\mathrm{#1}}}
\newcommand{\T}{{\mathrm{T}}}
\newcommand{\herm}{{\mathrm{H}}}

\newcommand{\h}{\mtxb{h}}
\newcommand{\x}{\mtxb{x}}
\newcommand{\R}{\mtxb{R}}
\newcommand{\W}{\mtxb{W}}
\newcommand{\w}{\mtxb{w}}

\newcommand{\uu}{\mtxb{u}}

\newcommand{\aRhof}{{\bm{{P}}}}

\newcommand{\Mset}{\mathcal{M}}

\newcommand{\Nset}{\mathcal{N}}

\title{Distributed Adaptive Norm Estimation For Blind System\\Identification in Wireless Sensor Networks}
\name{M. Blochberger\(^1\), F. Elvander\(^2\), R. Ali\(^1\), J. Østergaard\(^3\), J. Jensen\(^3\),  M. Moonen\(^1\), T. van Waterschoot\(^1\)\thanks{This research work was carried out at the ESAT Laboratory of KU Leuven, in the frame of the SOUNDS European Training Network. This project has received funding from the European Union's Horizon 2020 research and innovation programme under the Marie Skłodowska-Curie grant agreement No.\,956369. This research received funding in part from the Research Foundation - Flanders (FWO) grant 12ZD622N as well as from the European Union's Horizon 2020 research and innovation program / ERC Consolidator Grant: SONORA (No.\,773268). This paper reflects only the authors' views and the Union is not liable for any use that may be made of the contained information. Source code available at \url{https://github.com/SOUNDS-RESEARCH/icassp2023-adapt-dist-avg}}}
\address{\(^1\)KU Leuven, Dept. of Electrical Engineering (ESAT), STADIUS, 3001 Leuven, Belgium\\\(^2\)Aalto University, Dept. of Information and Communications Engineering, 02150 Espoo, Finland\\\(^3\)Aalborg University, Dept. of Electronic Systems, 9220 Aalborg, Denmark}

\begin{document}
\ninept
\maketitle
\begin{abstract}
    Distributed signal-processing algorithms in (wireless) sensor networks often aim to decentralize processing tasks to reduce communication cost and computational complexity or avoid reliance on a single device (i.e., fusion center) for processing.
    In this contribution, we extend a distributed adaptive algorithm for blind system identification that relies on the estimation of a stacked network-wide consensus vector at each node, the computation of which requires either broadcasting or relaying of node-specific values (i.e., local vector norms) to all other nodes.
    The extended algorithm employs a distributed-averaging-based scheme to estimate the network-wide consensus norm value by only using the local vector norm provided by neighboring sensor nodes.
    We introduce an adaptive mixing factor between instantaneous and recursive estimates of these norms for adaptivity in a time-varying system.
    Simulation results show that the extension provides estimation results close to the optimal fully-connected-network or broadcasting case while reducing inter-node transmission significantly.
\end{abstract}
\section{INTRODUCTION}
\label{sec:intro}

Distributed algorithms have been an active area of research for quite some time, with numerous control, optimization, and signal processing applications.
With the ever-growing number of smart multimedia devices in today's surroundings providing ubiquitous processing and communication capabilities, distributed audio and speech signal processing have also found their way into the spotlight.
Algorithms for distributed signal estimation \cite{5483092}, noise control and echo cancellation \cite{9670697}, as well as beamforming \cite{6663655,6329934,MARKOVICHGOLAN20154,6309434} to name a few, have been proposed.
Another essential task in audio and communication applications, which has received less attention, is multi-channel system identification, i.e., estimating channel responses in the time or frequency domain.
Distributed single-input-multiple-output (SIMO) blind system identification (BSI) has been addressed with adaptive cross-relation-based (CR) algorithms \cite{yuDistributedBlindSystem2014, liuDistributedBlindIdentification2016}.
In this context, we recently introduced an adaptive CR-based algorithm \cite{blochbergerDBSI} using the alternating direction method of multipliers (ADMM) \cite{boydDistributedOptimizationStatistical2011}.

The aforementioned distributed BSI algorithms rely on shared information between neighboring sensor nodes within the network.
However, the CR-based BSI task necessitates a non-triviality constraint on the full system to be identified (we refer the reader to, e.g., \cite{huangAdaptiveMultichannelLeast2002,huangClassFrequencydomainAdaptive2003} for details), which manifests itself as one or more variables that require network-wide shared information for their computation.
The algorithm in \cite{blochbergerDBSI} relies on node-wise values being relayed throughout the network.
In contrast, to overcome the need for the network to be fully connected, \cite{yuDistributedBlindSystem2014, liuDistributedBlindIdentification2016} use an average consensus \cite{xiaoFastLinearIterations2004} approach where a secondary recursion estimates the global variable for each signal frame.
Both approaches introduce additional transmissions of variables between nodes, the number of which, depending on network and neighborhood size, can be substantial or even unfeasible depending on the application.

In this paper, we extend our algorithm in \cite{blochbergerDBSI} with a distributed averaging-based \cite{xiaoFastLinearIterations2004} estimation scheme for the global variable.
This approach allows us, similarly to \cite{6334305,9914798}, to compute the global variable without needing a fully connected network or broadcasting.
We also introduce a mixing factor to include instantaneous values into the averaging recursion, which then allows us (i) to reduce the number of secondary iterations significantly (from 50 in existing approaches down to 1) and (ii) track time-varying systems.
Simulation results demonstrate how the proposed reduced-communication algorithm using in-neighborhood information delivers BSI performance close to an idealized one where network-wide information is available.

\section{DISTRIBUTED ADAPTIVE BSI IN SENSOR NETWORKS WITH ONLINE-ADMM}
\label{sec:dbsi}
In this section, we briefly outline the adaptive SIMO BSI algorithm proposed in \cite{blochbergerDBSI}, setting the scene for the distributed averaging scheme proposed herein.
We refer the reader to \cite{blochbergerDBSI} for the full derivation of the algorithm.

\subsection[]{Sensor network}
We assume a set of network nodes with one sensor each, that is, each node provides one signal.
Let this set of nodes be \(\Mset \triangleq \{1,\ldots,M\}\) and \(\mathcal{E}\) be a set of edge tuples, which connect the nodes forming a sensor network.
Each edge is an unordered pair of node indices \(\{i,j\} \in \mathcal{E}\), which represents the communication link, which we assume is instantaneoud and error free, between them.
We allow \(\{i,i\} \in \mathcal{E}\), in order to simplify notation later, however, this does not represent a link but rather indicates the obvious fact that node \(i\) has information about itself.
Furthermore, let \(\Nset_i = \{j|\{i,j\} \in \mathcal{E}\}\) denote the neighborhood of node \(i\).
We can define the symmetric adjacency matrix \(\mtxb{C}\) with elements \(C_{ij} = 1\) if \(\{i,j\} \in \mathcal{E}\) and 0 otherwise.
It may be noted that in \cite{blochbergerDBSI}, the pairs \(\{i,j\} \in \mathcal{E}\) are ordered, which yields a directed graph and leads to two sets of neighborhoods (``transmit'' and ``receive'') and therefore a non-symmetric adjacency matrix.
For the sake of brevity, we limit the explanations in this paper to the undirected case.
The case of directed graphs can be extended directly from the presented results.

\subsection[]{Signal model}
We consider a SIMO system with
\begin{align}
    \mtxb{s}(n) &= [s(n),\,\ldots,\,s(n-2L+2)]^{\T},\\
    \x_i(n) &= [x_i(n),\,\ldots,\,x_i(n-L+1)]^{\T}, \quad i \in \Mset,
\end{align}
the \(2L \times 1\) input signal frame and \(M\) \(L \times 1\)  output signal frames, respectively.
Each output \(\x_i(n)\) is the convolution of \(\mtxb{s}(n)\) with the respective channel impulse response \(\h_i\) and an additive noise term \(\mtxb{v}_i(n)\), assumed to be zero-mean and uncorrelated with \(\mtxb{s}(n)\).
The signal model is
\begin{equation}
    \x_i(n) = \mtxb{H}_i \mtxb{s}(n) + \mtxb{v}_i(n),
\end{equation}
with \(\mtxb{H}_i\), the \(L \times (2L-1)\) linear convolution matrix of the \(i\)th channel using the elements of \(\h_i\) of length \(L\).
For the purpose of this paper, the length \(L\) of the impulse responses is assumed to be known.

\subsection[]{Distributed BSI with Online-ADMM}
In BSI, the cross-relation problem formulation only uses output signals \(\x_i(n)\), exploiting relative information between them, to identify the system, i.e., the acoustic or communication channels \(\h = [\h_1^\T,\,...,\,\h_M^\T]^\T\).
The solution to this problem is found by the minimization problem \cite{langtongBlindIdentificationEqualization1994,huangAdaptiveMultichannelLeast2002,huangClassFrequencydomainAdaptive2003,blochbergerDBSI}:
\begin{equation}
    \begin{aligned}
        \hat{\h} = \arg \min_{\h} \quad &\h^\herm \R \h \\
        \text{s.t. } \quad &\|\h\| = 1,
    \end{aligned}\label{eq:frequency_domain:min_prob}
\end{equation}
where \(\hat{\h} = [\hat{\h}_1^\T,\,...,\,\hat{\h}_M^\T]^\T\) is the vector of estimated channel responses and \(\R\) is the so-called cross-relation (CR) matrix.
In \cite{blochbergerDBSI}, this problem is solved by a distributed adaptive algorithm that is based on separating the problem \eqref{eq:frequency_domain:min_prob} into node-wise subproblems,
\begin{equation}
    \begin{aligned}
        \underset{\w,\,\h}{\min} \quad &\sum_{i \in \Mset} \w_i^\herm \aRhof_i \w_i\\
        \text{s.t.} \quad &(\w_i)_j = \h_{\mathcal{G}(i,j)}\quad i \in \Mset,\,j \in \Nset_i\\
        &\|\h\| = 1.
    \end{aligned}\label{eq:general_consensus_admm:min_prob}
\end{equation}
Here, each subproblem is represented by a cost function \(\w_i^\herm \aRhof_i \w_i\), with the node-local channel estimates \(\w_i\) and CR matrix \(\aRhof_i\).
Both are analogous to \(\h_i\) and \(\R\) in \eqref{eq:frequency_domain:min_prob}, however, only using a subset of channels, corresponding to \(\Nset_i \subset \Mset\).
The first constraint enforces a consensus of local estimates, and \(\mathcal{G}(i,j)\) is a map equating local estimates of nodes connected by edge \((i,j) \in \mathcal{E}\).
The general-form consensus alternating direction method of multipliers (ADMM) \cite{boydDistributedOptimizationStatistical2011} is applied in an adaptive updating scheme (Online-ADMM) \cite{wangOnlineAlternatingDirection2013,hosseiniOnlineDistributedADMM2014}.
ADMM leads to update steps for local, consensus, and dual variables.
We refer the reader to \cite{blochbergerDBSI} for details.
The algorithm's consensus variable is the system's estimate, \(\hat{\h}\).
Each node \(i\) computes the respective subvector \(\hat{\h}_i\) locally, which is the computation step of interest for the extension proposed in this paper.
It is defined as
\begin{equation}
    \hat{\h}_i^{(n)} = \frac{\bar{\h}_i^{(n)}}{\sqrt{\sum_{j \in \Mset} \|\bar{\h}_j^{(n)}\|^2}}\label{eq:online_admm:consensus_update}
\end{equation}
for each node \(i \in \Mset\) with the local unnormalized consensus \(\bar{\h}_i^{(n)} = \bar{\w}_i^{(n)} + \frac{1}{\rho} \bar{\uu}_i^{(n-1)}\), a combination of in-neigborhood averages of node-wise local estimates \(\bar{\w}_i^{(n)}\) and dual (Lagrangian) variables \(\bar{\uu}_i^{(n-1)}\), where \(\rho\) is the ADMM penalty parameter.
A proper introduction of these variables is unfortunately out of scope for this paper, so for details, see \cite{blochbergerDBSI}.
The computation of the denominator of \eqref{eq:online_admm:consensus_update} requires  \(\|\bar{\h}_j^{(n)}\|\) from all nodes \(j \in \Mset\), which however, is not possible without a fully-connected network or broadcasting.
In \cite{blochbergerDBSI}, we assume that partial squared norms are relayed/broadcasted throughout the network until all nodes have the information necessary.

Subsequently, we introduce an extension to the algorithm described in \cite{blochbergerDBSI} using a fastest distributed linear averaging (FDLA) approach \cite{xiaoFastLinearIterations2004} to avoid the need for network-wide data transmission.
A related approach described in \cite{yuDistributedBlindSystem2014,liuDistributedBlindIdentification2016} employs a similar iterative averaging scheme within a distributed gradient descent algorithm for BSI in order to estimate a global variable.
It, however, necessitates many iterations per frame.
To alleviate this, we further introduce an adaptive mixing factor to include instantaneous values in the recursion to separate averaging iterations onto time frames of the adaptive algorithm.

\section{Adaptive norm estimation}
\label{sec:adaptivenormest}

\subsection[]{Distributed averaging}
First, assuming we have some unnormalized consensus variables \(\bar{\h}_i\) for \(i \in \Mset\), we look to distributively compute a sum of their squared norms, \(\sum_{j \in \Mset} \|\bar{\h}_j\|^2\), for \eqref{eq:online_admm:consensus_update} at each node \(i\).
Distributed linear combinations allow us to compute an average in a distributed manner, which makes computing the desired sum straightforward.
They generally have the form of
\begin{equation}
    \phi_i({k+1}) = \sum_{j \in \Nset_i} W_{ij} \phi_j({k}),\quad i\in \Mset,\label{eq:adaptivenormest:distlincomb}
\end{equation}
with iteration index \(k=0,...,K\).
This can be written in vector form \(\bm{\phi}({k+1}) = \W \bm{\phi}({k})\) with \(\bm{\phi}(k) = \begin{bmatrix} \phi_1(k) & \ldots & \phi_M(k) \end{bmatrix}^\T\).
We want to find the matrix \(\mtxb{W}\), which will ensure that for any initial value \(\bm{\phi}({0})\), each element of \(\bm{\phi}({k})\) converges to the average of the elements of \(\bm{\phi}({0})\) when \(k \to \infty\).
In this application, the elements of \(\bm{\phi}(0)\) are the initial node-wise squared norm values \(\|\bar{\h}_i\|^2\), which we want to compute the sum of.
This leads to \(\lim_{k \to \infty} M \phi_i(k) = \sum_{j \in \Mset} \|\bar{\h}_j\|^2\) for all \(i \in \Mset\), \cite{xiaoFastLinearIterations2004}.
Computing the desired \(\mtxb{W}\) is done by solving the fastest distributed linear averaging (FDLA) problem, introduced in \cite{xiaoFastLinearIterations2004}, in the form of
\begin{equation}
    \begin{aligned}
        \min& \quad \| \W - \bm{1}\bm{1}^\T/M\|_2\\
        \text{s.t.}& \quad \W \in \mathcal{C},\, \bm{1}^\T \W = \bm{1}^\T,\, \W \bm{1}= \bm{1}.
    \end{aligned}\label{eq:adaptivenormest:fdlaminprob}
\end{equation}
Here, the set \(\mathcal{C} = \{\W \in \mathbb{R}^{M \times M} \vert W_{ij} = 0 \text{ if } \{i,j\} \notin \mathcal{E}\}\) describes all matrices with the same sparsity pattern as the adjacency matrix \(\mtxb{C}\).
Now, the \(i\)-th row of \(\W\) contains the neigborhood weights for node \(i\) as non-zero entries.
It is to note that this has to be done only once for any fixed-topology network \(\Mset\) and \(\mathcal{E}\).
\subsection[]{Proposed adaptive estimation}
\begin{figure}[t]
    \centering
    \input{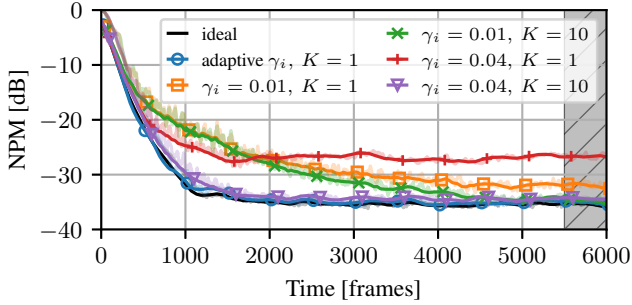}
    \vspace*{-0.6cm}
    \caption[]{Median NPM of 30 Monte-Carlo runs over time for ideal case where all node information is available (fully connected network/broadcasting), fixed values of \(\gamma_i\) with \(K \in \{1,10\}\), adaptive \(\gamma_i\) with \(K=1\). A moving average filter was applied to result curves for better readability (compare transparent and opaque lines). Hatched area indicates frames used in computations for \autoref{fig:simulations:avgNPMgamma}.}
    \label{fig:simulations:NPMtime}
\end{figure}
Given that the unnormalized consensus estimates \(\bar{\h}_i^{(n)}\) \cite{blochbergerDBSI} are time-varying, computing a distributed sum of their norms has to be performed adaptively as well.
The baseline approach is to compute \(K\) iterations \eqref{eq:adaptivenormest:distlincomb} per frame index \(n\), where \(K\) has to be sufficiently large for convergence (e.g., \(K=50\) in \cite{yuDistributedBlindSystem2014,liuDistributedBlindIdentification2016}).
This is undesirable, as each iteration requires that a node \(i\) shares the averaging variable \(\phi_i(k)\) with its neighborhood \(\Nset_i\).
Therefore, we first propose to split iterations over time frames.
One obvious approach would be - at each frame \(n\) - to set the initial values \(\phi_i^{(n)}(0)\) to the result of the \(K\)-th iteration \(\phi_i^{(n-1)}(K)\) of the previous frame \(n-1\).
With sufficiently large \(n\), convergence will be reached even when \(K\) is chosen small.
However, this rudimentary modification does not factor in the adaptive approach to channel identification.
That is, when letting \(\phi_i^{(0)}(0) = \|\bar{\h}_i^{(0)}\|^2\), with \(\bar{\h}_i^{(0)}\) being estimated at the inital frame \(n=0\), it converges to the sum of these initial - poor - estimates.
Instead, we factor in the instantaneous approximation, the weighted in-neighborhood sum \(\eta_i^{(n)} = \sum_{j \in \Nset_i} W_{ij} \|\bar{\h}_i^{(n)}\|^2\), which will introduce new information gained by more accurate estimates of \(\bar{\h}_i^{(n)}\).
We add this information by linearly combining it with the \(K\)-th iterate of \(\phi_i^{(n-1)}(k)\) and set it as the initial value fo the following frame
\vspace*{-0.1cm}
\begin{equation}
    \phi_i^{(n)}(0) = \gamma_i \eta_i^{(n)} + (1-\gamma_i) \phi_i^{(n-1)}(K),
\end{equation}
where \(0 \leq \gamma_i \leq 1\) is a mixing factor.
We exploit the instantaneous information \(\eta_i^{(n)}\) to find values for the mixing factor.
Experimental observations show that convergence is faster with larger \(\gamma_i\) while steady-state error is lower with smaller \(\gamma_i\).
We postulate that when the estimation of \(\bar{\h}_i\) is converging to a steady state, the absolute difference between subsequent frames \(| \eta_i^{(n)} - \eta_i^{(n-1)} | \to 0\).
Following from this, we say that after fast convergence with larger \(\gamma_i\), the algorithm's emphasis should then lie on norm estimation, i.e., the distributed averaging recursion.
This means \(\gamma_i \to 0\).
Based on this heuristic, we set \(\gamma_i^{(n)}\) proportional to the absolute difference of instantaneous approximations between subsequent frames and set an upper limit at 1,
\vspace*{-0.3cm}
\begin{equation}
    \gamma_i^{(n)} = \min \left\lbrace \frac{| \eta_i^{(n)} - \eta_i^{(n-1)} |}{\eta_i^{(n-1)}},\,1\right\rbrace.\label{eq:adaptivenormest:adaptivegamma}
\end{equation}
The initial value for \(\eta_i^{(n-1)}\) for \(n=0\) can be set to an arbitrary real number, in our case 1.

The adaptive estimate \(\phi_i^{(n)}(K)\) is then used in the denominator of \eqref{eq:online_admm:consensus_update} as \(\sqrt{M \, \phi_i^{(n)}(K)}\).
Refer to \autoref{alg:davg_norm_est} for the order of computations of the extended algorithm we introduced here.

\section{Evaluation}
\label{sec:simulations}
\begin{figure}[t]
    \centering
    \input{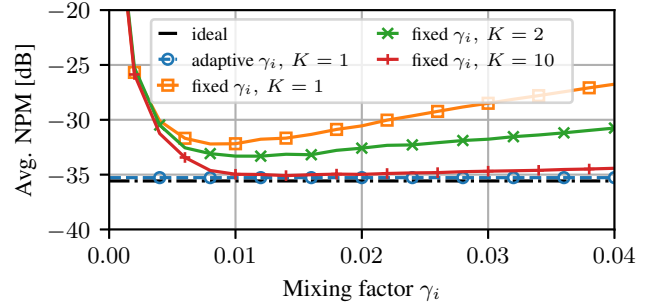}
    \vspace*{-0.6cm}
    \caption[]{Mean of 500 post-convergence frames of median NPM of 30 Monte-Carlo runs for fixed \(\gamma_i\) on the interval \([0.0, 0.04]\) and \(K \in \{1,2,10\}\). Results for ideal algorithm, and adaptive \(\gamma_i\) included for comparison.}
    \label{fig:simulations:avgNPMgamma}
\end{figure}
\subsection[]{Communication cost}
\label{sec:transcost}
To describe the communication cost within the network, we count the number of variables transmitted per time frame \(n\).
The most useful measure in the case of this work is to analyze the cost per node.
We compare the following communication schemes within the context of the algorithm \cite{blochbergerDBSI} this paper proposes the extension to:
\begin{itemize}
    \itemsep-0.2em
    \item[(1)] A fully-connected network, where node \(i\) communicates \(\|\bar{\h}_i\|^2\) to all other nodes \(\{j \in \Nset_i = \Mset\}\) directly, i.e., the neighborhood is the full network.
    \item[(2)] The node \(i\) communicates \(\|\bar{\h}_i\|^2\) only to neighboring nodes \(\{j \in \Nset_i \subset \Mset\}\) and applies \(K\) distributed averaging iterations.
\end{itemize}
\autoref{tab:transcost:table} compares the number of transmit and receive operations node \(i\) has to apply in order to compute \eqref{eq:online_admm:consensus_update}.
Further, it shows the complexity of the local problem at node \(i\), assuming it uses all information that it receives.
The advantage of the fully distributed algorithm becomes clear when \(M \gg N_i\), as the additional communication and local complexity depends on neighborhood size \(N_i\) and not on network size \(M\).
\vspace*{-0.6em}
\renewcommand{\arraystretch}{1.2}
\begin{table}[h]
    \centering
    \begin{tabular}{ |l|l|l|l| }
        \hline
        & Trans. Ops. & Rec. Ops. & Local Compl. \\
        \hline\hline
        (1) Fully con. & \(M-1\) & \(M-1\) & \(\mathcal{O}(M^2 L^2)\) \\
        \hline
        (2) Neighborh. & \(N_i K\) & \(N_i K\) & \(\mathcal{O}(N_i^2 L^2)\)\\ 
        \hline
    \end{tabular}
    \caption[]{Communication cost per node and frame \(n\) for computation of \eqref{eq:online_admm:consensus_update}.}
    \label{tab:transcost:table}
\end{table}
\renewcommand{\arraystretch}{1.0}

\subsection[]{Simulations}
To evaluate the effectiveness of the proposed extension, simulations were run, where we define the error measure as the normalized projection misalignment
\begin{equation}
        \text{NPM}(n) = 20\,\log_{10} \left(\left\| \h - \frac{\h^\T \hat{\h}^{(n)}}{\h^\T \h}\hat{\h}^{(n)} \right\| / \left\|\hat{\h}^{(n)}\right\|\right),
\end{equation}
commonly used in BSI (e.g., \cite{huangAdaptiveMultichannelLeast2002,huangClassFrequencydomainAdaptive2003,habetsOnlineQuasiNewtonAlgorithm2010}) to compare the estimated \(\hat{\h}^{(n)}\) and true \(\h\).

The first simulation setup is a network with \(M=5\) nodes arranged in a ring topology, each node having \(N_i=3\) neighbors (including node \(i\)).
The input signal is zero-mean white Gaussian noise (WGN), the impulse responses of length \(L=16\) are drawn from a normal distribution, and at each channel, independent WGN is added with \(\text{SNR}=10\,\text{dB}\).
The norms of the impulse responses are scaled to random values drawn from the uniform distribution \(\mathcal{U}_{[0.5,2.0]}\).
A comparison is made between the following cases:
\begin{itemize}
    \itemsep-0.2em
    \item[(a)] \(\|\bar{\h}_i\|^2\) of all \(i \in \Mset\) is available for global norm computation (``ideal'' in \autoref{fig:simulations:NPMtime} \& \autoref{fig:simulations:avgNPMgamma}),
    \item[(b)] Inter-neighborhood communication with fixed \(\gamma_i \in [0.0, 0.4]\) and \(K \in \{1,2,10\}\),
    \item[(c)] Inter-neighborhood communication with adaptive \(\gamma_i\) \eqref{eq:adaptivenormest:adaptivegamma} and \(K=1\).
\end{itemize}
\begin{algorithm}[t]
    \caption{ADMM BSI with distributed-averaging-based adaptive estimation of norm values}\label{alg:davg_norm_est}
    \(\W \gets\) \eqref{eq:adaptivenormest:fdlaminprob}\;
    \(\eta_i^{(n-1)} \gets 1, \forall i \in \Mset\)\;
    \For(){\(n=0\dots\)}
    {
        \For(){\(i \in \Mset\)}
        {
            \emph{The steps before are as introduced in }\cite{blochbergerDBSI}\\
            \dotfill\\
            \(\eta_i^{(n)} \gets \sum_{j \in \Nset_i} W_{ij} \|\bar{\h}_i^{(n)}\|^2\)\;
            \(\gamma_i^{(n)} \gets \min \left\lbrace \frac{| \eta_i^{(n)} - \eta_i^{(n-1)} |}{\eta_i^{(n-1)}},\,1\right\rbrace\)\;
            \eIf(){\(n = 0\)}
            {
                \(\phi_i^{(0)}(0) \gets \|\bar{\h}_i^{(0)}\|^2\)\;
            }
            {
                \(\phi_i^{(n)}(0) \gets \gamma_i^{(n)} \eta_i^{(n)} + (1-\gamma_i^{(n)}) \phi_i^{(n-1)}(K)\)\;
            }
            \For(){\(k=0,\dots,K\)}
            {
                Transmit \(\phi_i^{(n)}(k)\) to nodes  \(j \in \Nset_i\)\;
                \(\phi_i^{(n)}({k+1}) \gets \sum_{j \in \Nset_i} W_{ij} \phi_j^{(n)}({k})\)\;
            }
            \(\hat{\h_i}^{(n)} \gets \frac{\bar{\h}_i^{(n)}}{\sqrt{M \phi_i^{(n)}(K)}}\)\;
            \dotfill\\
            \emph{The steps after are as introduced in }\cite{blochbergerDBSI}\\
        }
    }
\end{algorithm}
\autoref{fig:simulations:NPMtime} shows the median of 30 Monte-Carlo runs.
We can observe that when fixing \(\gamma_i\), a tradeoff between convergence speed and steady-state error arises at low iteration counts (cf. also \autoref{fig:simulations:avgNPMgamma}).
With \(K=10\), the performance comes close to the optimal case at the cost of additional communication between nodes.
This in contrast to the case with adaptive \(\gamma_i\) where even at \(K=1\), i.e., no additional communication for the recursive estimation scheme, convergence speed and steady-state error are close to the optimal case.

\begin{figure}[t]
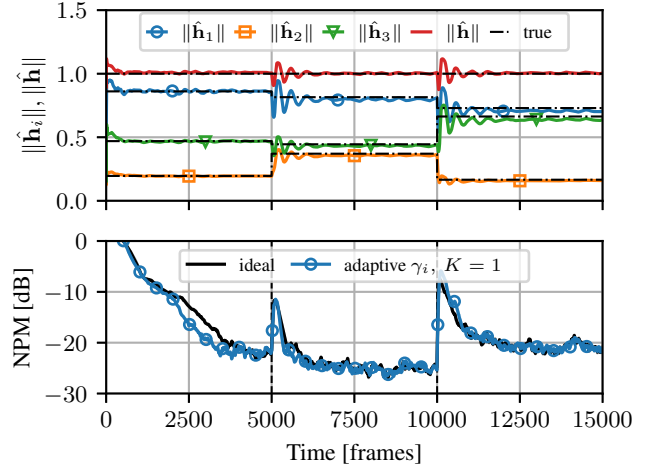

    \centering
    \input{simulations/plots/icassp2023-dynamic-norms/icassp2023-dynamic-norms.pgf}\\\vspace*{-1.2cm}
    \input{simulations/plots/icassp2023-dynamic-time/icassp2023-dynamic-time.pgf}
    \vspace*{-0.6cm}
    \caption[]{Results of 30 Monte-Carlo runs: Top: Estimated \(\|\h^{(n)}\|\), \(\|\hat{\h}_i^{(n)}\|\) over time, shown is the mean. Dot-dashed lines are true values. Bottom: Median NPM, comparing idealized algorithm and proposed extended algorithm. Dashed vertical lines indicate rescaling events.}
    \label{fig:simulations:NPMtimedyn}
\end{figure}

The second simulation setup is a \(3\)-node ring topology with a neighborhood size of \(N_i=2\) (including node \(i\)).
The input signal, impulse responses, additive noise, and SNR have the same parameters as the first simulation.
To test the algorithms response to time-varying scenarios, the 3 impulse responses are then scaled to norms of \(\{2.2, 0.5, 1.2\},\, \{2.2, 1.0, 1.2\},\, \{2.2, 0.5, 2.0\}\) at frames \(n \in \{0,\, 5000,\, 10000\}\) respectively.
\autoref{fig:simulations:NPMtimedyn} (bottom) shows the median NPM of 30 Monte-Carlo runs, where we can observe very similar behavior of the optimal and distributed-averaging-based algorithms, where both algorithms reach a converged state after the rescaling events.
It further shows the damped oscillations that appear in the estimates of \(\|\hat{\h}^{(n)}\|\) around the desired value 1 and \(\|\hat{\h}_i^{(n)}\|\) around the rescaled true norms in \autoref{fig:simulations:NPMtimedyn} (top), respectively.

In summary, the simulations show that depending on the choice of fixed \(\gamma_i\), it can lead to results close to an idealized case even with \(K=1\) distributed averaging iterations.
Further, introducing an adaptive \(\gamma_i^{(n)}\) allows the algorithm to converge faster and deal with time-variant systems.
This is an improvement over \cite{blochbergerDBSI} as inter-node communication is only required within node neighborhoods and over \cite{yuDistributedBlindSystem2014, liuDistributedBlindIdentification2016} needing only \(K=1\) instead of \(K=50\) iterations per frame.

\section[]{Conclusions}
\label{sec:conclusions}
In this contribution, we propose an extension to a distributed adaptive BSI algorithm applied in sensor networks.
Based on distributed averaging, only using the information provided by neighboring nodes in the sensor network, we compute estimates of norm values of channel estimates in order to enforce a norm constraint.
By balancing the averaging with introducing new data into the recursion, we allow the algorithm to follow an adaptive updating scheme.
The mixing factor, which balances new data with recursion data, is set adaptively dependent on instantaneous channel estimate norms.
These introductions reduce inter-node transmissions for each time frame while still delivering steady-state estimation performance close to the optimal case where network-wide information is available at all nodes.
The reduction that can be achieved goes as far as only needing a single iteration and, therefore, one additional inter-neighborhood information exchange per time frame.
We illustrate the performance in simulations with a white Gaussian input signal and random impulse responses drawn from the normal distribution.

\vfill\pagebreak

\bibliographystyle{IEEEbib}
\bibliography{bib_abbrev,refs}

\begin{thebibliography}{10}

\bibitem{5483092}
Alexander Bertrand and Marc Moonen,
\newblock ``Distributed adaptive node-specific signal estimation in fully
  connected sensor networks—part ii: Simultaneous and asynchronous node
  updating,''
\newblock {\em IEEE Trans. Signal Process.}, vol. 58, no. 10, pp. 5292--5306,
  2010.

\bibitem{9670697}
Santiago Ruiz, Toon van Waterschoot, and Marc Moonen,
\newblock ``Distributed combined acoustic echo cancellation and noise reduction
  in wireless acoustic sensor and actuator networks,''
\newblock {\em IEEE/ACM Trans. Audio Speech Lang. Process.}, vol. 30, pp.
  534--547, 2022.

\bibitem{6663655}
Yuan Zeng and Richard~C. Hendriks,
\newblock ``Distributed delay and sum beamformer for speech enhancement via
  randomized gossip,''
\newblock {\em IEEE/ACM Trans. Audio Speech Lang. Process.}, vol. 22, no. 1,
  pp. 260--273, 2014.

\bibitem{6329934}
Shmulik Markovich-Golan, Sharon Gannot, and Israel Cohen,
\newblock ``Distributed multiple constraints generalized sidelobe canceler for
  fully connected wireless acoustic sensor networks,''
\newblock {\em IEEE Trans. Audio Speech Lang. Process.}, vol. 21, no. 2, pp.
  343--356, 2013.

\bibitem{MARKOVICHGOLAN20154}
Shmulik {Markovich-Golan}, Alexander Bertrand, Marc Moonen, and Sharon Gannot,
\newblock ``Optimal distributed minimum-variance beamforming approaches for
  speech enhancement in wireless acoustic sensor networks,''
\newblock {\em Signal Processing}, vol. 107, pp. 4--20, 2015.

\bibitem{6309434}
Richard Heusdens, Guoqiang Zhang, Richard~C. Hendriks, Yuan Zeng, and
  W.~Bastiaan Kleijn,
\newblock ``Distributed mvdr beamforming for (wireless) microphone networks
  using message passing,''
\newblock in {\em Proc. 2012 Int. Workshop Acoustic Signal Enhancement (IWAENC
  '12)}, 2012, pp. 1--4.

\bibitem{yuDistributedBlindSystem2014}
Chengpu Yu, Lihua Xie, and Yeng~Chai Soh,
\newblock ``Distributed blind system identification in sensor networks,''
\newblock in {\em Proc. 2014 IEEE Int. Conf. Acoust., Speech, Signal Process.
  (ICASSP '14)}, May 2014, pp. 5065--5069.

\bibitem{liuDistributedBlindIdentification2016}
Ying Liu, Hao Liu, and Chunguang Li,
\newblock ``Distributed blind identification of sparse channels in sensor
  networks,''
\newblock in {\em Proc. 35th Chinese Control Conf.(CCC '16)}, July 2016, pp.
  5122--5127.

\bibitem{blochbergerDBSI}
Matthias Blochberger, Filip Elvander, Randall Ali, Marc Moonen, Toon van
  Waterschoot, Jan Østergaard, and Jesper Jensen,
\newblock ``Distributed cross-relation-based frequency-domain blind system
  identification using {{Online-ADMM}},''
\newblock in {\em Proc. 2022 Int. Workshop Acoustic Signal Enhancement (IWAENC
  '22)}, 2022, pp. 1--5.

\bibitem{boydDistributedOptimizationStatistical2011}
Stephen Boyd, Neal Parikh, Eric Chu, and Jonathan Eckstein,
\newblock ``Distributed {{Optimization}} and {{Statistical Learning}} via the
  {{Alternating Direction Method}} of {{Multipliers}},''
\newblock {\em Foundations and Trends in Machine Learning}, vol. 3, no. 1, pp.
  1--122, 2011.

\bibitem{huangAdaptiveMultichannelLeast2002}
Yiteng~Arden Huang and Jacob Benesty,
\newblock ``Adaptive multi-channel least mean square and {{Newton}} algorithms
  for blind channel identification,''
\newblock {\em Signal Processing}, p.~12, 2002.

\bibitem{huangClassFrequencydomainAdaptive2003}
Yiteng Huang and J.~Benesty,
\newblock ``A class of frequency-domain adaptive approaches to blind
  multichannel identification,''
\newblock {\em IEEE Trans. Signal Process.}, vol. 51, no. 1, pp. 11--24, Jan.
  2003.

\bibitem{xiaoFastLinearIterations2004}
Lin Xiao and Stephen Boyd,
\newblock ``Fast linear iterations for distributed averaging,''
\newblock {\em Syst. Control Lett.}, vol. 53, no. 1, pp. 65--78, Sept. 2004.

\bibitem{6334305}
Toon van Waterschoot and Marc Moonen,
\newblock ``Distributed estimation and equalization of room acoustics in a
  wireless acoustic sensor network,''
\newblock in {\em Proc. 20th European Signal Process. Conf. (EUSIPCO '12)},
  2012, pp. 2709--2713.

\bibitem{9914798}
Bilgesu Çakmak, Thomas Dietzen, Randall Ali, Patrick Naylor, and Toon~van
  Waterschoot,
\newblock ``A distributed steered response power approach to source
  localization in wireless acoustic sensor networks,''
\newblock in {\em Proc. 2022 Int. Workshop Acoustic Signal Enhancement (IWAENC
  '22)}, 2022, pp. 1--5.

\bibitem{langtongBlindIdentificationEqualization1994}
{Lang Tong}, {Guanghan Xu}, and T.~Kailath,
\newblock ``Blind identification and equalization based on second-order
  statistics: A time domain approach,''
\newblock {\em IEICE Trans. Inf. Theory}, vol. 40, no. 2, pp. 340--349, Mar.
  1994.

\bibitem{wangOnlineAlternatingDirection2013}
Huahua Wang and Arindam Banerjee,
\newblock ``Online {{Alternating Direction Method}} (longer version),''
\newblock {\em Proc. 29th Int. Conf. Machine Learning (ICML '12)}, July 2012.

\bibitem{hosseiniOnlineDistributedADMM2014}
Saghar Hosseini, Airlie Chapman, and Mehran Mesbahi,
\newblock ``Online distributed {{ADMM}} via dual averaging,''
\newblock in {\em Proc. 53rd IEEE Conf. Decision Control (CDC '14)}, Dec. 2014,
  pp. 904--909.

\bibitem{habetsOnlineQuasiNewtonAlgorithm2010}
Emanuel~A.P. Habets and Patrick~A. Naylor,
\newblock ``An online quasi-{{Newton}} algorithm for blind {{SIMO}}
  identification,''
\newblock in {\em Proc. 2010 IEEE Int. Conf. Acoust., Speech, Signal Process.
  (ICASSP '10)}, Mar. 2010, pp. 2662--2665.

\end{thebibliography}

\end{document}